\documentclass[a4paper,11pt]{article}
\usepackage{pos}

\newcommand{\mD}{\left(\begin{array}{cc} \DO & \Dd \\  \Ddb&\DOb \end{array} \right)}
\newcommand{\DO}{D_\Omega}
\newcommand{\Dd}{D_\partial}
\newcommand{\DOi}{D_\Omega^{-1}}

\newcommand{\DOb}{D_{\bar{\Omega}}}
\newcommand{\Ddb}{D_{\bar{\partial}}}
\newcommand{\DObi}{D_{\bar{\Omega}}^{-1}}

\newcommand{\Pdb}{\mathbb{P}_{\bar{\partial}}}

\newcommand{\Phidb}{\mathbb{\phi}_{\bar{\partial}}}

\newcommand{\hR}{\hat R}
\newcommand{\hDO}{\hat D_\Omega}
\newcommand{\hDd}{\hat D_\partial}
\newcommand{\hDOi}{\hat D_\Omega^{-1}}
\newcommand{\hPd} {\hat{\mathbb{P}}_\partial}

\newcommand{\hDOb}{\hat D_{\bar{\Omega}}}
\newcommand{\hDdb}{\hat D_{\bar{\partial}}}
\newcommand{\hDObi}{\hat D_{\bar{\Omega}}^{-1}}
\newcommand{\hPdb}{\hat{\mathbb{P}}_{\bar{\partial}}}

\newcommand{\mul}[1]{\left(\begin{array}{cc}#1 & 0 \\ 0& 0\end{array}\right)}
\newcommand{\mur}[1]{\left(\begin{array}{cc}0  & #1\\ 0& 0\end{array}\right)}
\newcommand{\mll}[1]{\left(\begin{array}{cc}0  & 0 \\ #1 & 0\end{array}\right)}
\newcommand{\mlr}[1]{\left(\begin{array}{cc}0  & 0 \\ 0& #1\end{array}\right)}

\newcommand{\mR}{\mul{ R}}
\newcommand{\mDO}{\mul{ \DO}}
\newcommand{\mDd}{\mur{ \Dd}}
\newcommand{\mDOi}{\mul{\DOi}}

\newcommand{\mDOb}{\mlr{\DOb}}
\newcommand{\mDdb}{\mll{\Ddb}}
\newcommand{\mDObi}{\mlr{\DObi}}
\newcommand{\mPdb}{\mul{\Pdb}}

\title{Algorithms for Domain Wall Fermions}
\ShortTitle{Algorithms for Domain Wall Fermions}

\author*[a,b]{Peter Boyle}
\author[c]{Dennis Bollweg}
\author[d]{Christopher Kelly}
\author[b]{Azusa Yamaguchi}

\affiliation[a]{Physics Department, Brookhaven National Laboratory, Upton, NY, USA}

\affiliation[b]{School of Physics and Astronomy, University of Edinburgh, Edinburgh UK}
\affiliation[c]{Physics Department, Columbia University, New York, USA}
\affiliation[d]{Computing Science Initiative, Brookhaven National Laboratory, Upton, NY, USA}

\emailAdd{pboyle@bnl.gov}
\emailAdd{ckelly@bnl.gov}
\emailAdd{db3516@columbia.edu}
\emailAdd{ayamaguc@staffmail.ed.ac.uk}

\abstract{We discuss algorithms for domain wall fermions focussing on accelerating Hybrid Monte Carlo sampling
  of gauge configurations. Firstly a new multigrid algorithm for domain wall solvers and secondly a
  Domain decomposed hybrid monte carlo approach applied to large subvolumes and optimised for GPU accelerated nodes.
  We propose a formulation of DD-RHMC that is suitable for the simulation of odd numbers of fermions. 
}

\FullConference{%
 The 38th International Symposium on Lattice Field Theory, LATTICE2021
  26th-30th July, 2021
  Zoom/Gather@Massachusetts Institute of Technology
}

\begin{document}
\maketitle

\section{Introduction}

Domain wall fermions are a theoretically attractive but numerically expensive formulation of lattice QCD.
In recent years valence analysis has become a largely solved problem with multiple propagator inversions
accelerated by deflation~\cite{Rudy,Shintani:2014vja,Clark:2017wom}
and able to run independently in a large system using MPI split communicators~\cite{GridManual}.
Current calculations use globally stored eigenvector deflation and solve
thousands of (deflated) propagators independently in parallel limiting communication overhead

In contrast, gauge configuration sampling is serially dependent and a strong scaling problem.
A single solution of the linear system is required for each quark mass and the serial dependence
of these requires scaling a single problem to as many nodes as possible.

Some recent systems such as those installed at Juelich and Edinburgh are well balanced
with communication and computation taking the same amount of time and delivering up to 10TF/s per node
on multiple node jobs with a volume per GPU in the region of $32^4$.

This balance will not be preserved in future systems, however.
GPU performance is likely to increase by as much as ten fold, while network performance may only increase
by 50\% in the next few years. In order to accelerate HMC evolution new algorithms will be required and in this
proceedings we document our plan to address this.

We plan to use a modified formulation of Domain Decomposed Hybrid Monte Carlo (DDHMC) and to combine this
with multigrid deflation of the local solves. The difference will be to focus on \emph{large} subdomains appropriate
to the entire volume processed by a large, multi-GPU computing node.

\section{Large volume domain decomposition}

We follow the DDHMC algorithm~\cite{Luscher:2004pav,DelDebbio:2006cn,DelDebbio:2007pz}
but aim to implement this rather differently.
Large blocks will be used, of $O(32^4)$ and no attempt will be made to keep the
Dirichlet boundary condition block operators well conditioned.
We aim to use this purely as a communication avoiding algorithm for multi-GPU nodes, rather than to precondition the HMC algorithm.
Prior implementations~\cite{DelDebbio:2006cn,DelDebbio:2007pz,PACS-CS:2009sof,Boku:2012zi}
on conventional multi-core microprocessors have found substantial computational acceleration
from smaller, cache resident cells processed by individual processor cores, reaching very high performance on
the Fugaku computer in particular~\cite{Kanamori:2021rwy,Ishikawa:2021iqw}.
As discussed by the original author of DDHMC, the locality factorisation of the determinant must
eventually win in computing systems with penalties for non-locality.

However, in highly parallel hardware optimising for cache locality is not possible as opportunities for sequential ordering of access is limited.
This proposed usage will keep the fraction of active links in the HMC at close to 100\%, while the force for the boundary determinant
will be suppressed by the width of the bands of inactive links. This may decorrelate better than the original DDHMC. The speed gain will come solely from avoiding communication slow down, and will reflect the nature and cost of computing and data access in modern supercomputers.

The large cell gives the opportunity for larger bands of inactive links than the original implementation, and peak forces may be suppressed
by the distance of propagation.

\subsection{Cuboidal domains and the Dirac operator}

We partition the lattice into two domains $\Omega$ and $\bar{\Omega}$.
Their \emph{exterior} boundaries
haloes are $\partial_\Omega$ and $\partial_{\bar{\Omega}}$ such that,
\begin{equation}\partial_\Omega \cap \Omega = \emptyset, \end{equation}
and
\begin{equation}\partial_{\bar{\Omega}} \cap \bar{\Omega} = \emptyset, \end{equation}
respectively.

The Dirac operator, with an appropriate non-lexicographic ordering
may then be written as
\begin{equation}
D = \mD,
\end{equation}
where $\Dd$ are terms in the matrix that connect the exterior boundary $\partial_\Omega$ to $\Omega$,
and $D_\Omega$ are terms in the matrix that connect $\Omega$ with itself, and similar for the other terms.
We write short hand,
\begin{equation} \begin{array}{cc}
\hDO = \mDO  &
\hDd = \mDd \\
\hDdb = \mDdb &
\hDOb = \mDOb 
\end{array}
\end{equation}
The Dirac operator may then be Schur factored as:
\begin{equation}
 \left(\begin{array}{cc} \DO & \Dd \\
                            \Ddb&\DOb \end{array}
   \right)
= 
\left(
\begin{array}{cc}
1  &  \Dd \DObi \\
0  & 1
\end{array}
\right)
\left(
\begin{array}{cc}
\DO - \Dd \DObi \Ddb & 0\\
0                    & \DOb
\end{array}
\right)
\left(
\begin{array}{cc}
1 &  0 \\
\DObi \Ddb  & 1
\end{array}
\right).
\end{equation}
We may then write the Fermion determinant as,
\begin{equation}
  \label{eq:schur}
\det D = \det{\DO} \det \DOb \det \left\{ 1 - \DOi \Dd \DObi \Ddb\right\},
\end{equation}
where we identify
\begin{equation}
\chi = 1 - \DOi \Dd \DObi \Ddb
\end{equation}
Following Luscher, we introduce projectors $\hPdb$ with both spinor and space structure
projecting all spinor elements in $\Omega$ connected by $\Ddb$ to $\bar{\Omega}$, and similarly $\hPd$.
The matrix $\hDdb$ acts only non-trivially on this subset of spinor components fields in $\partial_{\bar{\Omega}}$,
\begin{equation}
\hDdb \hPdb = \mDdb \mPdb  = \mDdb,
\end{equation}
and so we introduce the matrix,
\begin{eqnarray}
\hR = \mR &=&  \hPdb - \hPdb \hDOi \hDd \hDObi \hDdb\\ &=& \hPdb - \mPdb \mDOi \mDd \mDObi \mDdb.
\end{eqnarray}
Since in the right basis $\chi$ takes the form
\begin{equation}
\chi= \left( \begin{array}{cc} 1-X & 0\\ Y & 1 \end{array}\right)
\end{equation}
we see that,
\begin{equation}
\det \chi = \det R = \det (1-X).
\end{equation}
We may therefore treat the determinant of $\chi$ via a usual pseudofermion integral
\emph{only} over those fields in the space projected by $\Pdb$, and call these fields $\Phidb$.
When $R$ is taken as matrix \emph{from this subspace to itself}, it is non-singular with 
an inverse we can compute:
\begin{eqnarray}
\hR^{-1} &=& \hPdb - \hPdb D^{-1} \hDdb\\ &=& \mPdb - \mPdb D^{-1} \mDdb.
\end{eqnarray}
This is most easily seen by inserting UDL decomposition of $D^{-1}$, showing that
$\hR \hR^{-1} = \hPdb$.

\subsection{Domain shapes}

\begin{figure}[hbt]
\includegraphics[width=0.5\textwidth]{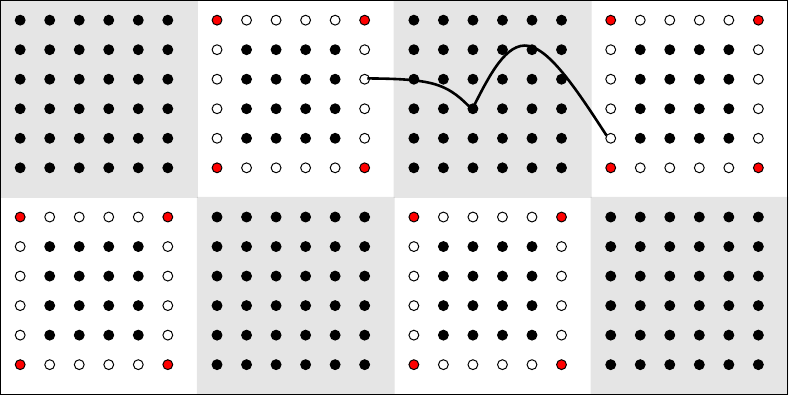}  
\includegraphics[width=0.3\textwidth]{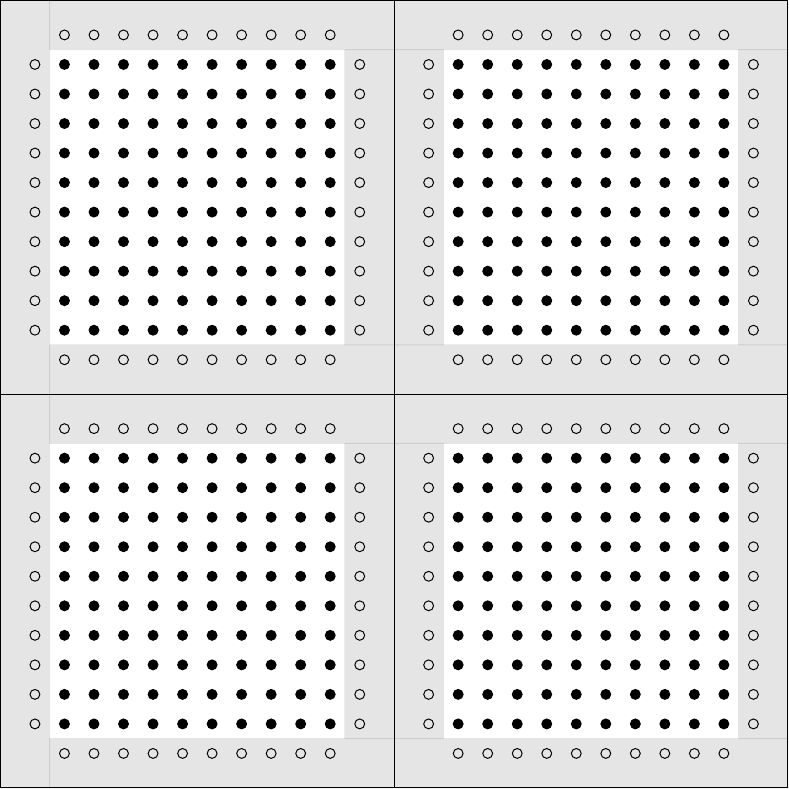}  
\caption{\label{domains}
Left:  Original small domain DDMHC domain layout.
    Sites containing spin projected pseudofermion support are labelled with open circles while
    sites with all spin components in pseudofermions are labelled filled red.
    The force for the boundary determinant is suppressed by two factors of the quark propagator. By using large domains
    we have control over the level of suppression and can reduce this significantly using inactive links while retaining a
   good fraction of active links in the evolution.
Right: alternate domain decomposition adopted in this work: domain $\bar \Omega$ is the union
    of all surface lattice sites on each computing node and is entirely inactive, while domain $\Omega$
    is interior. The forces in $\Omega$ are somewhat reduced.
 }
\end{figure}

The original algorithm selected $\Omega$ and $\bar \Omega$ to be sites located on even and odd
cells with a hyper-cuboidal decomposition, Figure \ref{domains}.
We seek to make the domains as large as possible. However, with a symmetrical block size for $\Omega$ and $\bar \Omega$,
one cannot increase the size of a domain to be close to an entire computing node. If the sub-domain were sized to a single
computing node, then sequences involving only $\hDOi$, for example, would not load balance on the machine and leave half the nodes
idle.

\begin{figure}[hbt]
\includegraphics[width=0.49\textwidth]{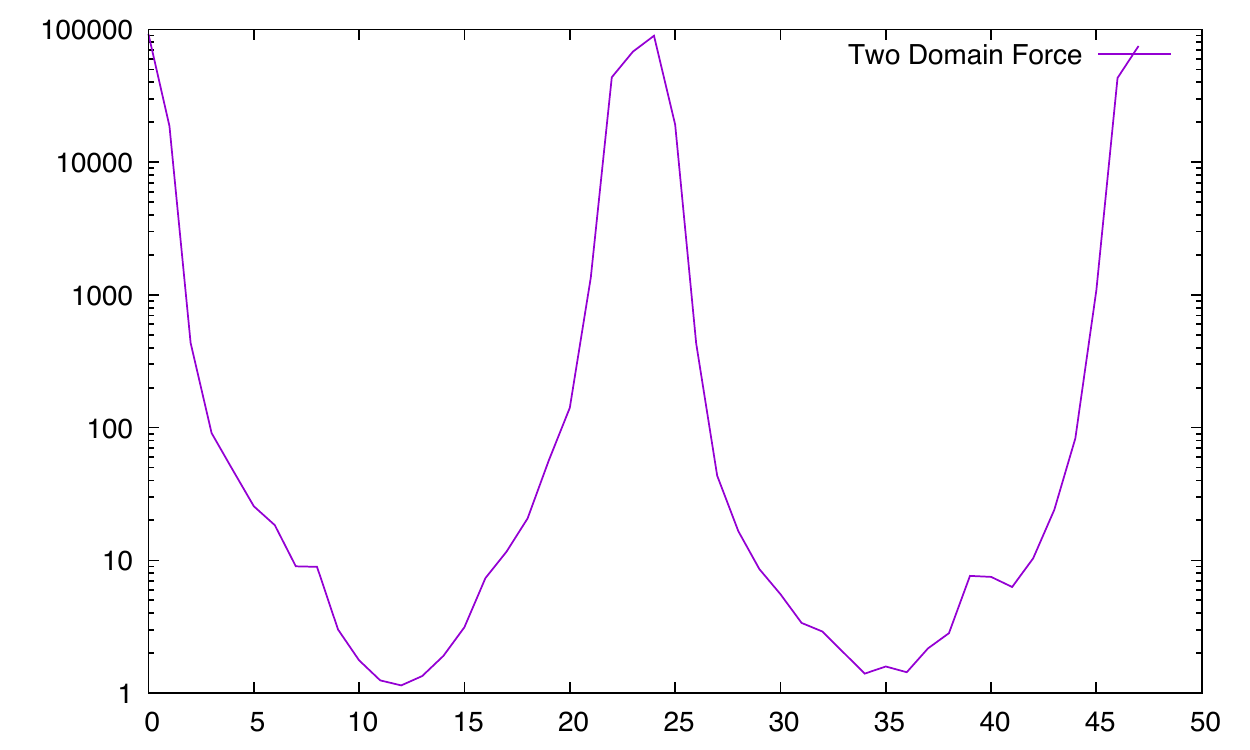} 
  \caption{\label{force} 
    Force (norm squared) profile from the boundary determinant in a one dimensional domain decomposition on a $16^3\times 48$ lattice.
    The hierarchy of force is visible and may be exploited. 
  }
\end{figure}

Therefore to maximise the number of active links we need a different domain decomposition scheme.
Communication avoidance suggests to make regions inactive only in the neighbourhood
of the boundaries between nodes, with the size of the inactive region controlled to suppress molecular dynamics
forces by however much is required to maintain a good suppression of communication.
This encouraged us to adopt a non-standard domain pattern, Figure \ref{domains}, where $\Omega$ is the interior
(non-boundary) cells of each node, while $\bar \Omega$ is union of all lattice sites on the
surfaces of all computing nodes. The entire domain $\bar \Omega$ is left inactive in the HMC evolution, along with
a controllable subset of links within $\Omega$. This allows minimises the forces,
 Figure~\ref{force}, in $\Omega$ and makes evolution of $\bar \Omega$ irrelevant. 

\subsection{Domain wall pseudofermion structure}

The local two flavour determinants are formed in the usual way from Hasenbusch ratios up to the Pauli
Villars mass. These solves can be performed independently on each node and decouple communication on the
machine. Within these, the usual red-black preconditioned HMC can be used. The domain $\bar \Omega$ has frozen
link variables throughout a HMC trajectory has no change in determinant. Full sampling is restored with
a random translation applied between trajectories.

The two flavour boundary pseudofermion action takes the form
\begin{equation}
\Phidb^\dag (R R^\dagger )^{-1} \Phidb .
\end{equation}
Where the boundary links are all inactive (so $\Ddb$ is \textit{not} differentiated) the force can be calculated as follows. The force term for the local determinant factor is standard but restricted to the local cell, and only active links. However all cells add together so the code implementation will be the standard one, and only the solver will differ.

For the pseudofermion derivative terms involving $R$, we have,
\begin{equation}
\delta R^{-1} = \Pdb D^{-1} \delta D  D^{-1} \Ddb. 
\end{equation}
The force is suppressed by quark propagation by the distance from the gauge link to the surface or plane of the domain boundary.
This may further be suppressed if more gauge links are kept inactive, perhaps a band of some depth around the plane connecting subvolumes.
This is a tunable parameter that pretty much guarantees we can obtain a reasonable ratio in the ``size'' of gauge forces.

A Hasenbusch ratio between matrices with differing mass $R$ and $P$ (e.g. Pauli Villars boundary) pseudofermion is formed with action,
\begin{equation}
\Phidb^\dag  P^\dagger R^{-\dagger}R^{-1} P   \Phidb.
\end{equation}
and we may differentiate this using
\begin{equation}
\delta R^{-1} = \Pdb D^{-1} \delta D  D^{-1} \Ddb. 
\end{equation}
and
\begin{equation}
\delta R =   \Pdb \DOi (\delta \DO ) \DOi \Dd \DObi \Ddb
           + \Pdb \DOi  \Dd \DObi (\delta \DOb) \DObi \Ddb .
\end{equation}
The pseudofermion derivative term therefore involves four PV local solves, and two non-local light quark solves.

\subsubsection{DD-RHMC}

We may introduce a large simplification and enable the simulation of single flavours
with a DD-RHMC approach as follows. We may rearrange and implement the boundary determinant
with a full, four-volume pseudofermion as:

\begin{equation}
\label{eq:ddrhmc}
\det \left\{ 1 - \DOi \Dd \DObi \Ddb\right\} = \frac{\det D}{\det{\DO} \det \DOb} = \frac{\det D}{\det D_{\rm dirichlet}}
\end{equation}

This RHS term may then be simulated with a standard RHMC one flavour ratio, similar to a Hasenbusch mass preconditioning
ratio pseudofermion, where the pseudofermion action takes the form
\begin{equation}
S_{1f}^{\rm boundary} = \phi^\dagger (D_{\rm dirichlet}^\dagger D_{\rm dirichlet})^{\frac{1}{4}}
                           (D^\dagger D)^{-\frac{1}{2}} (D_{\rm dirichlet}^\dagger D_{\rm dirichlet})^{\frac{1}{4}}
\phi
\end{equation}
The local Dirichlet determinants can be inserted, including Hasenbusch mass ratio terms using either
a Rational HMC pseudofermion action or an Exact One Flavour Action (EOFA) approach.

\subsection{Results }

We carried out an initial implementation of DDHMC by introducing a wrapper for all Grid~\cite{Boyle:2016lbp}
Fermion action objects that applies Dirichlet boundary conditions to its gauge field argument.
The communications may be shut off for local domain solves.
The present implemention has a restriction that the suppressed communication boundaries
must align with GPU boundaries, however it is planned to generalise the code for multi-GPU nodes to align the domain
boundaries at whole node boundaries.

The code was designed with flexibility and the domain shapes are in principle completely general.
A projector to the domain is implemented and edges are detected with a mask and shift by one approach in each
direction. This allows for algorithmic flexibility and rapid prototyping at the expense of lower efficiency in
a performance non-critical part of the code.

A standard two flavour Pseudofermion ratio object is sufficient to simulate the local determinant factors,
while a new domain decomposed two flavour ratio object was introduced for the boundary determinant.

\begin{figure}[hbt]
\includegraphics[width=0.5\textwidth]{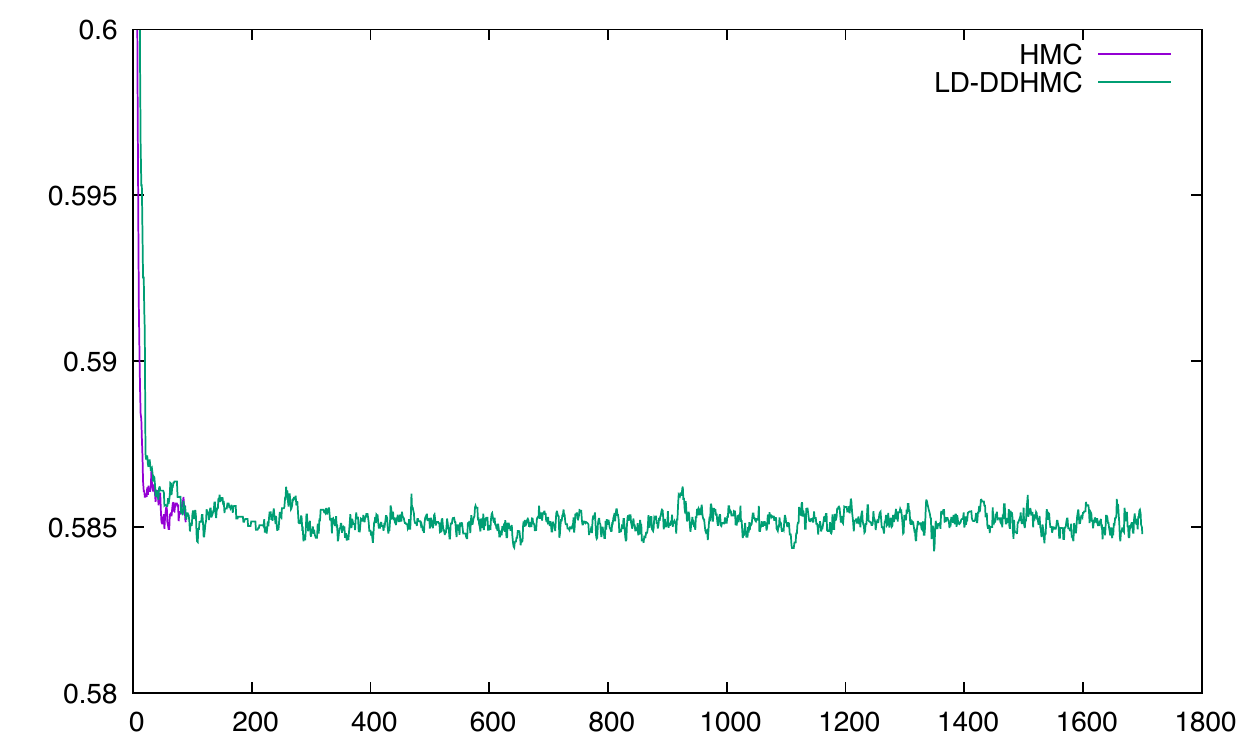}  
\caption{\label{plaq}
  Plaquette log for two flavour DWF simulation on $16^3 \times 48$ with $m_f=0.01$ and Iwasaki gauge action at $\beta=2.13$.
  The HMC and DDHMC agree validating the implementation.
}
\end{figure}

We have evolved a two flavour $16^3 \times 48$ system with the Iwasaki gauge action and $\beta = 2.13$.
This was subdivided into GPUs, of size $16^3\times 24$ each and the interior cells of $16^3\times 22$ evolved
in each trajectory. The plaquette time history is displayed in Figure~\ref{plaq}.

During this evolution a timestep ratio of 4:1 between the boundary determinant and the local determinant was
maintained with only three steps per trajectory in the boundary determinant and 12 for the local determinant
using the Omelyan integrator.

\begin{figure}[hbt]
\includegraphics[width=0.5\textwidth]{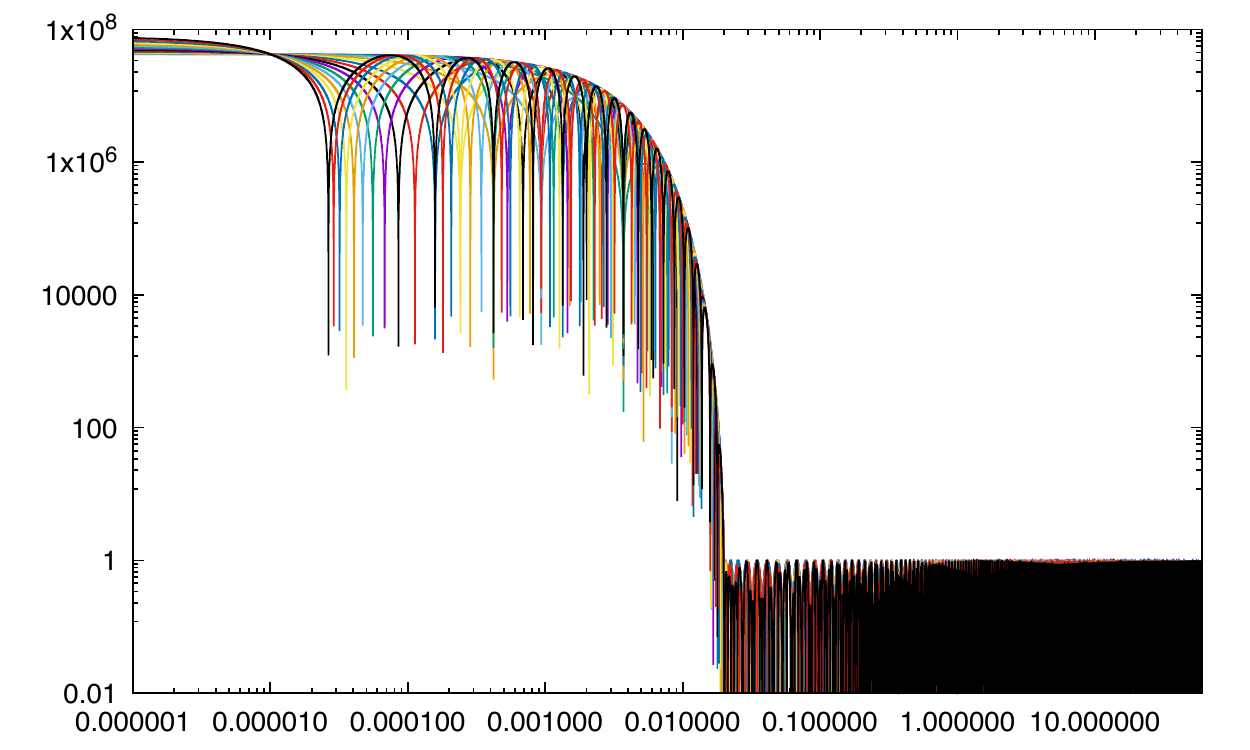}
\caption{\label{cheby}
  Filter functions obtained with Chebyshev filters for fast set up multigrid. These have been demonstrated
  to \emph{both} set up and solve a multigrid algorithm faster than a single standard red-black preconditioned Krylov solver for the
  Shamir DWF case. This is appropriate to use for local domain solves inside DDHMC evolution.
}
\end{figure}

We have prototyped a DDHMC algorithm in the Grid library.
It runs efficiently on GPU computing nodes, supporting CUDA, HIP, and SYCL for all large US supercomputer
architectures. Multi-core CPUs are also supported but these are not presently the target of the software
optimisation and algorithmic tuning.
Initial results suggest correct implementation and a substantially reduced level of communication is enabled.

The DD-RHMC approach has been proposed in this proceedings and used to add strange quarks with a mass of $a m_s=0.04$ on this
ensemble. Preliminary testing suggests the HMC forces are similarly well controlled, even with the four-volume pseudofermion
approach to the boundary determinant. This generalises the domain decomposition approach to odd flavours.

Although multigrid for Domain Wall Fermions is less mature~\cite{Cohen:2011ivh,Boyle:2014rwa,Brower:2020xmc,Boyle:2021wcf}
than for Wilson fermions~\cite{Luscher:2007se,Brannick:2007ue,Babich:2010qb}
we intend to combine the evolution with a recently developed approach to set up multigrid algorithms
quickly using Chebyshev filters inside the HMC algorithm. We believe the filter functions, Figure~\ref{cheby},
(which can be obtained recursively) offer good subspace generation with lower cost than
inverse iteration~\cite{Boyle:2021wcf}.

\section{Acknowledgements}

PB has been supported in part by the U.S. Department of Energy, Office of Science, Office of Nuclear Physics under the Contract No. DE-SC-0012704 (BNL).
A.Y. and D.B. have been supported by Intel.


\begin{thebibliography}{99}

\bibitem{Rudy}
``Non-Perturbative Renormalization and Low Mode Averaging with Domain Wall Fermions'',  R. Arthur, PhD Thesis, University of Edinburgh.

  
\bibitem{Shintani:2014vja}
E.~Shintani, R.~Arthur, T.~Blum, T.~Izubuchi, C.~Jung and C.~Lehner,
``Covariant approximation averaging,''
Phys. Rev. D \textbf{91} (2015) no.11, 114511
doi:10.1103/PhysRevD.91.114511
[arXiv:1402.0244 [hep-lat]].
%98 citations counted in INSPIRE as of 29 Nov 2021

\bibitem{Clark:2017wom}
M.~A.~Clark, C.~Jung and C.~Lehner,
``Multi-Grid Lanczos,''
EPJ Web Conf. \textbf{175} (2018), 14023
doi:10.1051/epjconf/201817514023
[arXiv:1710.06884 [hep-lat]].
%20 citations counted in INSPIRE as of 29 Nov 2021


\bibitem{GridManual}
``Grid Documentation''
  https://github.com/paboyle/Grid/blob/develop/documentation/Grid.pdf
  
\bibitem{Luscher:2004pav}
M.~Luscher,
``Schwarz-preconditioned HMC algorithm for two-flavour lattice QCD,''
Comput. Phys. Commun. \textbf{165} (2005), 199-220
doi:10.1016/j.cpc.2004.10.004
[arXiv:hep-lat/0409106 [hep-lat]].
%252 citations counted in INSPIRE as of 29 Nov 2021

\bibitem{DelDebbio:2006cn}
L.~Del Debbio, L.~Giusti, M.~Luscher, R.~Petronzio and N.~Tantalo,
``QCD with light Wilson quarks on fine lattices (I): First experiences and physics results,''
JHEP \textbf{02} (2007), 056
doi:10.1088/1126-6708/2007/02/056
[arXiv:hep-lat/0610059 [hep-lat]].
%148 citations counted in INSPIRE as of 29 Nov 2021


\bibitem{DelDebbio:2007pz}
L.~Del Debbio, L.~Giusti, M.~Luscher, R.~Petronzio and N.~Tantalo,
``QCD with light Wilson quarks on fine lattices. II. DD-HMC simulations and data analysis,''
JHEP \textbf{02} (2007), 082
doi:10.1088/1126-6708/2007/02/082
[arXiv:hep-lat/0701009 [hep-lat]].
%101 citations counted in INSPIRE as of 29 Nov 2021



\bibitem{PACS-CS:2009sof}
S.~Aoki \textit{et al.} [PACS-CS],
``Physical Point Simulation in 2+1 Flavor Lattice QCD,''
Phys. Rev. D \textbf{81} (2010), 074503
doi:10.1103/PhysRevD.81.074503
[arXiv:0911.2561 [hep-lat]].
%184 citations counted in INSPIRE as of 29 Nov 2021


\bibitem{Boku:2012zi}
T.~Boku, K.~I.~Ishikawa, Y.~Kuramashi, K.~Minami, Y.~Nakamura, F.~Shoji, D.~Takahashi, M.~Terai, A.~Ukawa and T.~Yoshie,
``Multi-block/multi-core SSOR preconditioner for the QCD quark solver for K computer,''
PoS \textbf{LATTICE2012} (2012), 188
doi:10.22323/1.164.0188
[arXiv:1210.7398 [hep-lat]].
%13 citations counted in INSPIRE as of 29 Nov 2021

\bibitem{Kanamori:2021rwy}
I.~Kanamori, K.~I.~Ishikawa and H.~Matsufuru,
``Object-oriented implementation of algebraic multi-grid solver for lattice QCD on SIMD architectures and GPU clusters,''
doi:10.1007/978-3-030-86976-2\_15
[arXiv:2111.05012 [hep-lat]].
%0 citations counted in INSPIRE as of 29 Nov 2021

\bibitem{Ishikawa:2021iqw}
K.~I.~Ishikawa, I.~Kanamori, H.~Matsufuru, I.~Miyoshi, Y.~Mukai, Y.~Nakamura, K.~Nitadori and M.~Tsuji,
``102 PFLOPS Lattice QCD quark solver on Fugaku,''
[arXiv:2109.10687 [hep-lat]].
%1 citations counted in INSPIRE as of 29 Nov 2021


\bibitem{Boyle:2016lbp}
P.~A.~Boyle, G.~Cossu, A.~Yamaguchi and A.~Portelli,
``Grid: A next generation data parallel C++ QCD library,''
PoS \textbf{LATTICE2015} (2016), 023
doi:10.22323/1.251.0023
%28 citations counted in INSPIRE as of 29 Nov 2021  


\bibitem{Luscher:2007se}
M.~Luscher,
``Local coherence and deflation of the low quark modes in lattice QCD,''
JHEP \textbf{07} (2007), 081
doi:10.1088/1126-6708/2007/07/081
[arXiv:0706.2298 [hep-lat]].
%150 citations counted in INSPIRE as of 29 Nov 2021

\bibitem{Brannick:2007ue}
J.~Brannick, R.~C.~Brower, M.~A.~Clark, J.~C.~Osborn and C.~Rebbi,
``Adaptive Multigrid Algorithm for Lattice QCD,''
Phys. Rev. Lett. \textbf{100} (2008), 041601
doi:10.1103/PhysRevLett.100.041601
[arXiv:0707.4018 [hep-lat]].
%55 citations counted in INSPIRE as of 29 Nov 2021


\bibitem{Babich:2010qb}
R.~Babich, J.~Brannick, R.~C.~Brower, M.~A.~Clark, T.~A.~Manteuffel, S.~F.~McCormick, J.~C.~Osborn and C.~Rebbi,
``Adaptive multigrid algorithm for the lattice Wilson-Dirac operator,''
Phys. Rev. Lett. \textbf{105} (2010), 201602
doi:10.1103/PhysRevLett.105.201602
[arXiv:1005.3043 [hep-lat]].
%106 citations counted in INSPIRE as of 29 Nov 2021

%\cite{Cohen:2011ivh}
\bibitem{Cohen:2011ivh}
S.~D.~Cohen, R.~C.~Brower, M.~A.~Clark and J.~C.~Osborn,
%``Multigrid Algorithms for Domain-Wall Fermions,''
PoS \textbf{LATTICE2011} (2011), 030
doi:10.22323/1.139.0030
[arXiv:1205.2933 [hep-lat]].
%17 citations counted in INSPIRE as of 29 Nov 2021

\bibitem{Boyle:2014rwa}
P.~A.~Boyle,
``Hierarchically deflated conjugate gradient,''
[arXiv:1402.2585 [hep-lat]].
%29 citations counted in INSPIRE as of 29 Nov 2021

\bibitem{Brower:2020xmc}
R.~C.~Brower, M.~A.~Clark, D.~Howarth and E.~S.~Weinberg,
``Multigrid for chiral lattice fermions: Domain wall,''
Phys. Rev. D \textbf{102} (2020) no.9, 094517
doi:10.1103/PhysRevD.102.094517
[arXiv:2004.07732 [hep-lat]].
%3 citations counted in INSPIRE as of 29 Nov 2021
  
%\cite{Boyle:2021wcf}
\bibitem{Boyle:2021wcf}
P.~Boyle and A.~Yamaguchi,
``Comparison of Domain Wall Fermion Multigrid Methods,''
[arXiv:2103.05034 [hep-lat]].
%0 citations counted in INSPIRE as of 29 Nov 2021


\end{thebibliography}
\end{document}